\documentstyle [12pt,a4p,epsfig,amsmath,multicol]{article}
\newcommand {\rbf}  {\overline{r}_f}
\newcommand {\vbf}  {\overline{v}_f}
\newcommand {\abf}  {\overline{a}_f}
\newcommand {\sbf}  {\overline{s}_f}
\newcommand {\sbq}  {\overline{s}_q}
\newcommand {\vbq}  {\overline{v}_q}
\newcommand {\abq}  {\overline{a}_q}
\newcommand {\rbQ}  {\overline{r}_Q}

\newcommand {\rbl}  {\overline{r}_l}
\newcommand {\vbl}  {\overline{v}_l}
\newcommand {\abl}  {\overline{a}_l}
\newcommand {\sbl}  {\overline{s}_l}
\newcommand {\rbc}  {\overline{r}_c}
\newcommand {\vbc}  {\overline{v}_c}
\newcommand {\abc}  {\overline{a}_c}
\newcommand {\sbc}  {\overline{s}_c}
\newcommand {\rbb}  {\overline{r}_b}
\newcommand {\vbb}  {\overline{v}_b}
\newcommand {\abb}  {\overline{a}_b}
\newcommand {\sbb}  {\overline{s}_b}
\newcommand {\gbl}  {\overline{g}_b^L}
\newcommand {\gbr}  {\overline{g}_b^R}
\newcommand {\gcl}  {\overline{g}_c^L}
\newcommand {\gcr}  {\overline{g}_c^R}
\newcommand {\Afbf}  {A_{FB}^{0,f}}

\newcommand {\tpol}  {$\tau$-polarisation}
\newcommand {\alps}  {\alpha_s(M_Z)} 
\begin{document}
\hspace*{8cm} {UGVA-DPNC 1998/07-177 July 1998}
\newline
 {\it Revised and corrected version of UGVA-DPNC 1997/10-172, CERN OPEN 97-033,
 hep-ph/9801355.To be published in Modern Physics Letters A.}
\vspace*{0.6cm}
\begin{center} 
{\bf INDICATIONS FOR AN ANOMALOUS RIGHT-HANDED
COUPLING OF THE b-QUARK FROM A MODEL INDEPENDENT ANALYSIS
OF LEP AND SLD DATA ON Z DECAYS}
\end{center}
\vspace*{0.6cm}
\centerline{\footnotesize J.H.Field}
\baselineskip=13pt
\centerline{\footnotesize\it D\'{e}partement de Physique Nucl\'{e}aire et 
 Corpusculaire, Universit\'{e} de Gen\`{e}ve}
\baselineskip=12pt
\centerline{\footnotesize\it 24, quai Ernest-Ansermet CH-1211Gen\`{e}ve 4. }
\centerline{\footnotesize E-mail: john.field@cern.ch}
\baselineskip=13pt
 
\vspace*{0.9cm}
\abstract{ A model independent analysis is made using the LEP and SLD data
on Z decays available at the end of 1996. The effective weak coupling
constants of leptons, c quarks and b quarks are extracted. Except for the
right-handed b quark coupling, they all agree with the predictions
of the Standard Electroweak Model for $m_t = 180$ GeV and $m_H = 100$ GeV.
The right-handed b quark coupling is found to be 42$\%$ and 3.2 standard
deviations above the Standard Model prediction.}
\vspace*{0.9cm}
\normalsize\baselineskip=15pt
\setcounter{footnote}{0}
\renewcommand{\thefootnote}{\alph{footnote}}
\newline 
{\it Keywords ;} Standard Electroweak Model, LEP and SLD data, Z-decays,
 Anomalous right-handed b quark coupling.
\newline

\vspace*{0.4cm}

\section{Introduction}
 This letter describes a model independent analysis based on a recent
compilation of LEP and SLD results on Z-decays~\cite{x1}. In the first stage
of the analysis the effective weak coupling constants of the
charged leptons, c quarks
and b quarks are extracted from the data using only weak theoretical
assumptions. In the second stage a
detailed comparison of the extracted effective couplings with the
 Standard Model (SM)~\cite{x2} predictions is
made. Confidence Levels (CLs) for consistency of the data with the
model are calculated, taking carefully into account all important error
correlations. Finally, the possible physical significance of the deviations
observed from the the SM
predictions for the b quark couplings is discussed.

\section{Extraction of the Effective Weak Coupling Constants}
Instead of the vector ($\vbf$) and axial vector ($\abf$) effective coupling
constants for a fermion (lepton or quark) $f$, the equivalent quantities
$\rbf$ and $\sbf$ are first extracted from the data. These are defined as:
 \begin{eqnarray}
 \rbf &\equiv &\vbf/\abf \\
 \sbf &\equiv &(\abf)^2+(\vbf)^2
 \end{eqnarray}
The experimental errors on $\rbf$ and $\sbf$ (unlike those on $\vbf$ and
$\abf$) are essentially uncorrelated, much simplifying the calculation
of CLs for consistency with theoretical predictions. In the first comparisons
with the SM shown below, the predictions are given by the global SM fit,
with $m_t = 172$ GeV and $m_H = 149$ GeV, from Ref.[1]. The value of the
t-quark mass found in this fit is in very good agreement with the 
directly measured value~\cite{x1} of  $m_t = 175(6)$ GeV \footnote
{Throughout this letter
 total experimental errors ( which, unless otherwise stated,
 are the quadratic sum of statistical and systematic errors)
 are given in terms of the last
 significant figures. 0.1533(27) denotes 0.1533$\pm$ 0.0027}. 
 The effect of varying the Higgs boson mass, the only remaining unknown 
 parameter of the SM, is later taken into account in the detailed
 comparisons with the extracted effective coupling constants.
 \par The quantities $\rbf$ ($f=$l,c,b) and $\sbl$ (l is a generic 
 lepton label) may be extracted directly from the data assuming only
 lepton universality. A further assumption is necessary in order to
 extract $\sbc$ or $\sbb$, and hence the c and b quark couplings. Since the
 couplings of the u,d,s quarks are only poorly measured~\cite{x3}, 
 `non-b quark lepton universality' is assumed. That is that $e$, $\mu$, $\tau$,
 u, d, s and c  are all assigned the same effective weak mixing angle.
 Another possiblity is to assume an independently measured value of
 $\alps$. The c and b quark couplings can then be extracted from $\rbc$,
 $\rbb$ and the ratios $\Gamma_Q/\Gamma_l$ ( $Q =$ c, b), independently of 
 the light quark effective couplings. The disadvantage of this method is that the
  extracted values of the heavy quark couplings are strongly correlated 
  with the assumed value of $\alps$.     
\par At LEP, $\rbf$ is found from the measured, corrected, pole forward/backward
charge asymmetries $\Afbf$~\cite{x1} via the relations:
  \begin{eqnarray}
  \Afbf & = & \frac{3}{4} A_e A_f     \\
  A_f &\equiv & \frac{2 \vbf \abf}{(\abf)^2+(\vbf)^2}= \frac{2 \rbf}{1+\rbf^2}
  \end{eqnarray}
  $A_e$ and $A_{\tau}$ have also been measured at LEP via the angular
   dependence of the $\tau$-polarisation asymmetry:
   \begin{equation}
   \overline{P}_{\tau}(\cos \theta) = -\frac{A_{\tau}+A_e F(\theta)}
   {1+A_{\tau} A_e F(\theta)}
   \end{equation}
   where 
   \[ F(\theta) \equiv 2 \cos \theta /(1+ \cos^2 \theta) \]
   and $\theta$ is the angle between the incoming $e^-$ and the outgoing 
   $\tau^-$ in the $\tau$-pair centre of mass frame.
   At SLD, $A_e$ is directly measured by the left/right beam polarisation 
   asymmetry $A_{LR}$, while $A_c$ and $A_b$ are determined from the
   left/right-forward/backward asymmetries of tagged heavy quarks.
   \par The separate LEP and SLD average values of the 
   electroweak observables, which are 
   directly sensitive to the effective couplings, are reported in 
   Table 1. Also shown in Table 1 are the total errors defined as the 
   quadratic sum of the statistical and systematic components
 ( $\sigma_{TOT} = \sqrt{\sigma_{STAT}^2+\sigma_{SYS}^2}$) as well as the
   systematic components $\sigma_{SYS}$. The SM predictions for 
   each electroweak observable and the deviations from the predictions 
   in units of $\sigma_{TOT}$ are also reported in Table 1. 
  The combined LEP/SLD averages of $A_l$, $A_c$, $A_b$, $R_c$
   and $R_b$, where $R_Q= \Gamma_Q/\Gamma_{had}$ , ( $Q =$ c, b) are reported
   in Table 2. 
       \par The values of $\rbf$ ($f=$l,c,b) derived from the measurements
   of $A_f$, using Eqn.(4), are presented in Table 3. 
 For the b quark, mass effects
     were taken into account by using the corrected form of Eqn.(4):
    \begin{equation}
    A_b = \frac{2 (\sqrt{1-4 \mu_b}) \rbb}{1-4 \mu_b+(1+2 \mu_b) \rbb^2}
    \end{equation} 
    where $\mu_b = (\overline{m}_b(M_Z)/M_Z)^2 \simeq 1.0 \times 10^{-3}$. The
    running b quark mass is taken as $\overline{m}_b(M_Z) = 3.0$ GeV~\cite{x4}.
    Agreement is seen with the SM at the $2 \sigma$ level for $\rbl$, at
     $< 1 \sigma$ for $\rbc$, but only at the $3.3 \sigma$ level for $\rbb$.
     The similar discrepancy for $A_b$ was mentioned, but not discussed
     in terms of the b quark couplings, in Ref.[1].
     \par It is of interest to study the sensitivity of the observed 
     deviation of $\rbb$ from the SM prediction to the treatment of the
     experimental errors. The deviation of $3.3 \sigma$ is found when the
  statistical and systematic errors for each electroweak observable are 
  added in quadrature. As estimates of systematic errors are often 
  conservative (i.e. too high), an upper limit on the possible size of
  the deviation is given by neglecting all systematic errors. On the other
  hand, it is not clear that systematic errors should necessarily be added
  quadratically to statistical ones, and also the meaning of a systematic
  error in terms of probability content is, usually, not clearly defined.
  An estimate of the lower limit of the deviation is given by adding 
  linearly the statistical and systematic errors for each electroweak
  observable in Table 1. Following these procedures, the following results
  are found:
   \[ \rbb = 0.586(27),~~3.8\sigma~~\rm{deviation}.~~
 (\sigma_{TOT}=\sigma_{STAT})               \]
\[ \rbb = 0.581(44),~~2.5\sigma~~\rm{deviaton}.~~ 
(\sigma_{TOT}=\sigma_{STAT}+\sigma_{SYS})     \]   
 It can be seen that the observed deviation remains at $\ge$ 2.5$\sigma$
 independently of the treatment of systematic errors.     
     \par The quantity $\sbl$ is derived from the leptonic width $\Gamma_l$
     using the relation:
     \begin{equation}
     \sbl = (\abl)^2+(\vbl)^2 = \frac{12 \pi \Gamma_l}{\sqrt{2} G_{\mu} M_Z^3}
     \frac{1}{(1+\frac{3 \alpha(M_Z)}{4 \pi})}
 \end{equation}
 The value obtained for $\sbl$, quoted in the first column of Table 4, uses the
 LEP average value of $\Gamma_l$ from Table 1 together with:
 $G_{\mu} = 1.16639 \times 10^{-5}$ (GeV)$^2$~\cite{x5}, $M_Z = 91.1863$ GeV, 
 and $\alpha(M_Z)^{-1} = 128.896$~\cite{x1}. Good agreement is found with the
 SM value. Solving Eqns.(1) and (2) for $\abl$ and $\vbl$ yields the results
 presented in Table 5. As in the calculation of all
 the other effective couplings, the signs of $\abl$ and $\vbl$ are chosen to be
 the same as the SM predictions. The values of $\abl$ and $\vbl$ are in
 good agreement with the LEP+SLD averages quoted in Ref.[1], taking into account
 the slightly different analysis procedures\footnote{Ref.[1] included small
 mass corrections in calculating $\abl$ and $\vbl$ which are neglected here.}.
 \par The quantities $\overline{s}_Q$ ( $Q =$ c,b), including quark mass
  effects, may be derived from the measured quantities $R_Q$ via the
  relation:  
  \begin{equation}
 \overline{s}_Q = (\overline{a}_Q)^2(1-6 \mu_Q)+(\overline{v}_Q)^2=
 \frac{R_Q S_Q}{(1-R_Q)C_Q^{QED}C_Q^{QCD}} 
 \end{equation}
 where
 \[ S_Q \equiv \sum_{q \ne Q}[(\overline{a}_q)^2(1-6 \mu_q)+(\overline{v}_q)^2]\]
 and~\cite{x6}:
 \begin{eqnarray}
 C_Q^i & = & 1+\delta_Q^i - < \delta_{q \ne Q}^i >~~~~~i = QED, QCD;~~~~ \mu_q
 = 0~~{\rm for}~~ q \ne b \nonumber \\
  \delta_q^{QED} & = & \frac{3 (e_q)^2}{4 \pi} \alpha (M_Z),~~
  \delta_{q \ne b}^{QCD} = 1.00a_s+1.42a_s^2,~~~
  \delta_b^{QCD} = .99a_s-1.55a_s^2  \nonumber
  \end{eqnarray}
  $q$ is a generic quark flavour index, $e_q$ the quark electric charge in
  units of that of the positron
  and $a_s \equiv \alps/\pi$ .
   $<X>$ denotes the quark flavour average of $X$.
   As mentioned above, $\mu_b = 1.0 \times 10^{-3}$
  while, taking into account the present experimental error on $R_c$, $\mu_c$
  is set to zero. The numerical values of the QED and QCD correction factors,
  with $\alpha_s(M_Z) = 0.12$ and $\alpha(M_Z)^{-1} = 128.9$, are presented in 
  Table 6. The non-b quark couplings in Eqn.(8) are written, conventionally,
   as:
  \begin{eqnarray}
   \overline{a}_q & = & \sqrt{\rho_q} ~T^q_3  \\
   \overline{v}_q & = & \sqrt{\rho_q} (T^q_3-2 e_q (\overline{s}^q_W)^2)
   \end{eqnarray} 
   where, assuming non-b quark lepton universality
   \footnote{ Here the weak isospin symmetry of the SM is invoked to
   calculate the unmeasured couplings. It is also assumed that the quantum
   corrections contained in $\rho_q$ and $(\overline{s}^q_W)^2$, though
   not necessarily those of the SM, are universal.}:
  \begin{eqnarray}
   \sqrt{\rho_q} & = & \sqrt{\rho_l}~ =  ~2 |\abl |~~~(~{\rm all}~ q \ne b~) \\
   (\overline{s}^q_W)^2 & = & \frac{1}{4}(1-\rbl)~~~~~~~(~{\rm all}~ q \ne b~)
 \end{eqnarray}
and $T_3^q$ is the third component of the weak isospin of the quark q. 
 Substituting the measured values of $\rbl$,  $\abl$, from Tables 3 and 5 and
 of $R_c$, $R_b$ from Table 2, leads to the values of $\sbc$, $\sbb$ reported
  in Table 4. Note that the value of $\sbb$, and hence $\abb$ and $\vbb$ are
  extracted first. The latter are then substituted into Eqn.(8) (taking into
  account their experimental errors) in order to find $\sbc$.
In Table 4 good agreement is seen between the measured values of 
$\sbl$ and $\sbc$ and the SM predictions. On the other hand, $\sbb$ lies 
1.3$\sigma$ above the prediction: a residual of
the well known `$R_b$ problem'~\cite{x1}.   
  Solving  Eqns.(1) and (8) then gives the
   effective coupling constants for the heavy quarks
   presented in Table 7. The values 
  found, as well as the errors, agree well with those reported by Renton
  in a recent review~\cite{x7}.
The solutions for $\abf$, $\vbf$ obtained from the essentially uncorrelated
quantities $\rbf$ and $\sbf$ are shown graphically in Figs1a,1b,1c for 
 f $=$ $l$,c,b respectively. 
It is clear from Fig.1c that largest discrepancy with the SM is in the
parameter $\rbb$(completely determined by $A_b$) rather than in $\sbb$
(~essentially determined by $R_b$). Indeed, if the SM value for the latter
is used, instead of the measured one, to solve for $\abf$ and $\vbf$, the 
discrepancies between the values found and the SM are almost unchanged.
\begin{figure}[htbp]
\begin{center}\hspace*{-0.5cm}\mbox{
\epsfysize10.0cm\epsffile{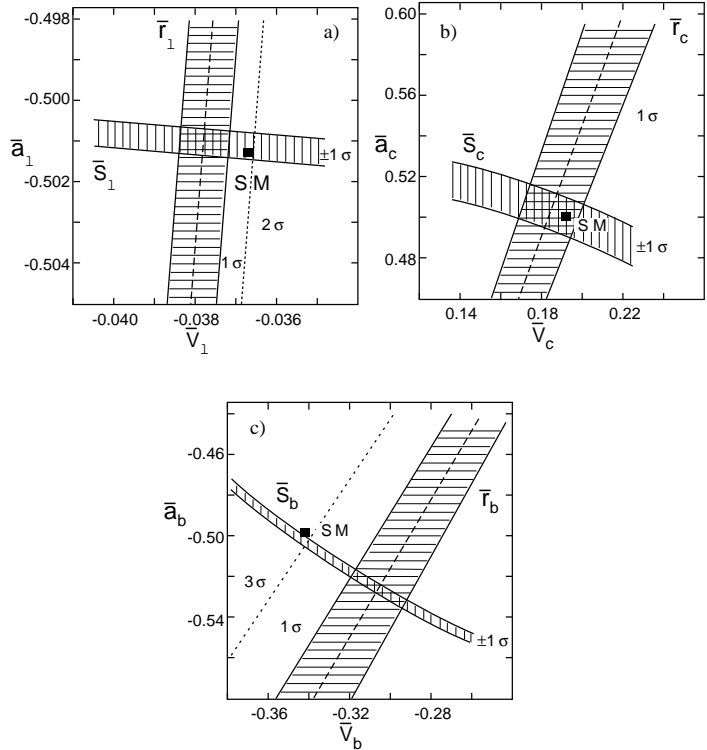}}
\caption{Constraints on the effective couplings 
$\abf$, $\vbf$ provided by the measurements of $\rbf$ and $\sbf$ .
a) leptons, b) c quarks, c) b quarks. The cross-hatched areas show
$\pm 1 \sigma$ limits. The dotted lines in a),[c)] show $2 \sigma$, 
[$3 \sigma$] limits for $\rbl$,[$\rbb$].
 SM is the Standard Model prediction for $m_t =$ 172 GeV, $m_H =$ 149 GeV.}
\label{fig-fig1}
\end{center}
 \end{figure}  
 The deviations in the b quark couplings found here are briefly compared
 with those that led, in 1995, to the `$R_b$ problem' at the end of this
 letter.  
  Although the c quark couplings agree well with the SM, and are
  also consistent with the quark-lepton
  universality hypothesis, both $\abb$ and $\vbb$ differ from the SM values
  by more than three standard deviations. The errors on these quantities are,
  however, highly correlated.The statistical significance of these deviations
  is discussed in detail below.
  \par It should be remarked that, although a particular value (0.12) of
   $\alpha_s(M_Z)$
  has been assumed in order to extract the effective couplings of the heavy
  quarks, the sensitivity to the chosen value is very weak. Varying
  $\alpha_s(M_Z)$ over the range $0.1 <  \alpha_s(M_Z) < 0.14$ leads 
  variations of only $\simeq 3 \times 10^{-4}$ in $\abb$ and $\vbb$ to be
   compared with experimental errors $\simeq 1-4 \times 10^{-2}$ (see Table 7).
\par A further constraint on the quark couplings is provided by the
 measurement of the mean forward/backward quark charge asymmetry:
 \begin{equation}
 \langle A_{FB}^q \rangle = \frac{8 A_l \sum_{q} \vbq \abq}{\sum_{q}[(1-6 \mu_q)
 (\abq)^2+(\vbq)^2]}
 \end{equation}
  All experimental analyses performed to date have assumed the correctness 
  of the SM and have used measurements of $\langle A_{FB}^q \rangle$ to
  determine a value of $\sin^2 \theta^{lept}_{eff}$~\cite{x1}. Inserting the
  average value of the latter reported in Ref.[1] into the SM formula
  for $\langle A_{FB}^q \rangle$ and propagating the error leads to the
  `measured' value:
  \[ \langle A_{FB}^q \rangle = 0.1592(86)  \]
  As shown in Table 8 this value is consistent with the SM prediction, 
  with the `model independent' prediction given by inserting the lepton and
 b quark couplings from Tables 5 and 7 into Eqn(13)
and assuming non-b quark lepton universality for the u,d,s,c quarks, as well as the
   prediction when, in the latter case, the measured b quark couplings are
   replaced by the SM ones. With the present experimental errors, 
   $\langle A_{FB}^q \rangle$ is therefore insensitive to possible deviations
   of the b quark couplings from the SM, of the magnitude observed in the
   $A_b$ measurements. 
\par As mentioned earlier, in order to avoid having to introduce an accurate 
value of $\alps$ as a correlated parameter in the extraction of the
heavy quark effective 
couplings, the hypothesis of non-b quark lepton universality was made in 
deriving the value of $\sbb$ from the measured quantity $R_b$. The consistency
of this assumption is now checked by extracting $\alps$ from the LEP average value
of $R_l \equiv \Gamma_{had}/\Gamma_l$~\cite{x1}:
\[ R_l = 20.778(29)    \]
using the relation:
\begin{equation}
R_l= 3 \frac{<C^{QED}_q> <C^{QCD}_q>}{C^{QED}_l}\frac{\sum_{q} \sbq}{\sbl}.
\end{equation}
The QED and QCD correction factors $<C^{QED}_q>$ and $<C^{QCD}_q>$
  are averaged over all quark flavours.
The QED correction factors are:
\[ <C^{QED}_q> = 1.00040,~~~~~C^{QED}_l = 1.00063  \]
Inserting the measured values of $\sbb$ and $\sbl$,and using non-b quark lepton
universality to evaluate $\sbq$ ( q $\ne$ b )
gives, for the QCD correction factor:
 \[<C^{QCD}_q>  = 1.0394(21)    \]
 Using the third order perturbative QCD formula~\cite{x8}:
 \begin{equation} 
 <C^{QCD}_q>  = 1+1.06 \frac{\alps}{\pi}+0.9 (\frac{\alps}{\pi})^2
 -15(\frac{\alps}{\pi})^3
 \end{equation} 
 gives:
 \[   \alps = 0.116_{-0.007}^{+0.005}  \]
 which may be compared to the global fit value of Ref.[1]:
 \[   \alps = 0.120(3)  \]
 The good agreement of the  model independent analysis  result with the
 global world average value: $\alps = 0.118(5)$ found in two recent reviews
 ~\cite{x9,x10} of all published measurements of $\alpha_s$, shows that
 an analysis assuming this value of $\alps$, but without the assumption of
 non-b quark lepton universality, would lead to essentially the same values
 of the b quark couplings as those reported in Table 7. In the fit used in Ref.[7]
 to determine the heavy quark effective couplings the constraint
 $\alps = 0.123(6)$ was imposed. As mentioned above, the fitted heavy quark couplings
 are very consistent with those found in the present analysis. 
    
 \par In order to correctly calculate the statistical significance of the
 deviations from the SM predictions of the effective couplings shown in 
 Tables 5 and 7 it is necessary to take into account the correlations between
 the errors on the different quantities. To avoid the very large correlations
 between the errors on $\abf$ and $\vbf$ (for the case of b quarks the 
 correlation coefficient is -0.96) it is convenient to use, in calculating the
 $\chi^2$, the equivalent quantities $\rbf$, $\sbf$ for which the
 errors are uncorrelated
 for a given fermion flavour $f$. Important correlations still exist, however, 
 between the errors on ($\rbl$, $\rbc$) and ($\rbl$, $\rbb$) in the case
 that $\rbc$ and $\rbb$ are extracted from forward/backward asymmetries
 using Eqns.(3),(4) and (6). The correlation coefficient is:
 \begin{equation}
 {\cal C}_{lQ} = -\frac{(1-\rbl^2)(1+\rbQ^2)}{(1+\rbl^2)(1-\rbQ^2)}
 \frac{\sigma_{\rbl}}{\rbl}\frac{\rbQ}{\sigma_{\rbQ}},~~~(Q=c,b) 
 \end{equation}
 Substituting the parameters from Table 3 gives:
 \[ {\cal C}_{lc} = -0.29,~~~~ {\cal C}_{lb} = -0.52 \]
 The results on the CLs for the agreement with the SM
 of different sets of effective
weak coupling constants, parameterised in terms of $\rbf$ and $\sbf$,
 are collected in Table 9.
 The CLs assume perfect statistical consistency of the different
 measurements contributing to the averages. 
The entries in the first column
 of Table 9, giving the level of agreement of ($\rbl$, $\sbl$) with the SM 
 prediction are simply calculated from the entries of Tables 3 and 4
 using a diagonal error matrix, since the errors on $\rbl$ and $\sbl$
 are uncorrelated.  
 Calculating separately the contributions to $\chi^2$ from $\rbl$ and $\rbb$,
 where the latter is derived from the LEP $A_{FB}^{0,b}$ measurement, and
 $\rbb$ derived via Eqn.(6) directly from the SLD $A_b$ measurement, 
 gives the entries reported in the second column of Table 9.  The CL for
 agreement with the SM prediction is 1.4\%. The third column 
 of Table 9 results from adding to the
 $\chi^2$ in the second column the (uncorrelated)\footnote{Actually there
 is a weak correlation between $\sbb$ and $\rbl$
 following from Eqn.(8), where $\rbl$ is used to calculate $S_Q$. However
 the correlation coefficient is only $\simeq  0.08$ and is neglected here.}
 contributions of $\sbl$ and $\sbb$. In the fourth column of Table 9 the 
 $\chi^2$ and CL of the variables $\rbl$, $\rbb$ and $\rbc$ taking into
 account the $\rbl$-$\rbb$ and $\rbl$-$\rbc$ correlations are given. In the
 last column of Table 9 the (uncorrelated) variables $\sbl$, $\sbb$ and $\sbc$
 are added to those of the fourth column. Note that the number of degrees
 of freedom corresponding to the $\chi^2$ values reported in the
second, third, fourth 
 and fifth columns of Table 9 are 3, 5, 5 and 8 respectively, since the 
 $\rbc$ and $\rbb$ measurements derived from the SLD $A_c$, $A_b$ determinations
 give separate, uncorrelated, contributions to the $\chi^2$. 
   As expected, the agreement with the SM improves as the number of degrees
  of freedom of the $\chi^2$ increases (the more parameters are considered,
  the more likely is a deviation associated with any of the parameters
  to be consistent with a statistical fluctuation).
\begin{figure}[htbp]
\begin{center}\hspace*{-0.5cm}\mbox{
\epsfysize10.0cm\epsffile{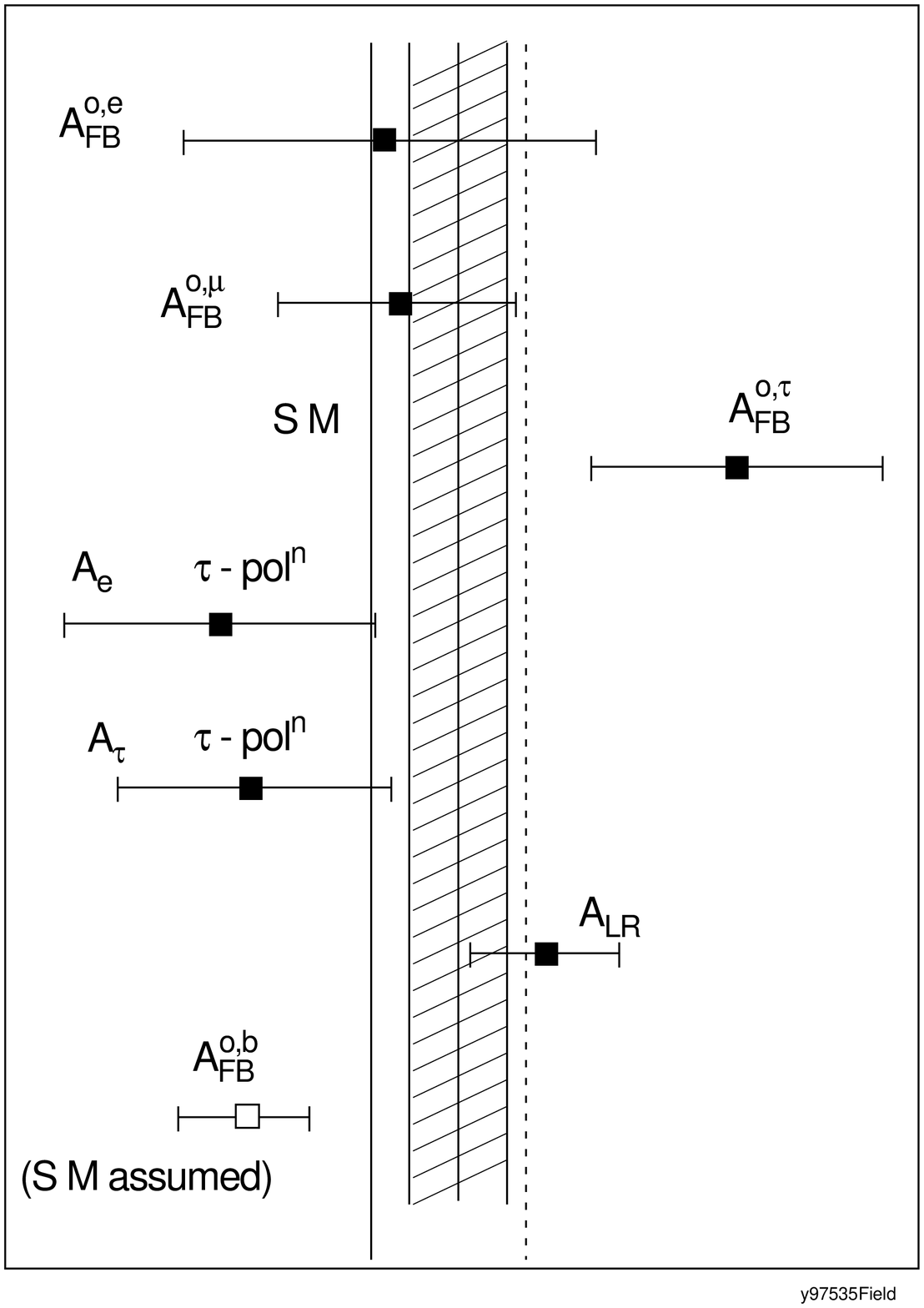}}
\caption{ LEP and SLD $A_l$ measurements. The hatched band shows
the $\pm 1\sigma$ region around the weighted average value. The weighted 
average value, excluding the \tpol~ measurements, is given by the 
dashed line. The solid line is the 
Standard Model prediction for
 $m_t =$ 172 GeV, $m_H =$ 149 GeV. The open square shows the value of
 $A_l$ derived from the LEP average value of $A_{FB}^{0,b}$ assuming the SM;
 this datum is not included in the weighted averages shown.}
\label{fig-fig2}
\end{center}
 \end{figure} 
  \par As previously stated, the CLs shown in Table 9
   are calculated on the assumption that the 
  different experiments contributing to the average values of $A_l$, $A_c$
  and $A_b$ are consistent with each other. The average value of $A_l$
  is also derived assuming lepton universality. Although the consistency
  of the nine different measurements of $A_{FB}^{0,b}$~\cite{x1} contributing 
  to the LEP average value of $A_b$ ~\cite{x1} with their weighted
 mean value is very good 
  ($\chi^2 = 5.9$ for 8 $dof$, CL$ = 66\%$) this is not the case for
  the different LEP and SLD measurements of $A_l$ shown in Fig. 2. The 
  overall consistency of the different measurements is only fair
  ($\chi^2 = 9.7$ for 5 $dof$, CL$ = 8.4\%$) and, as discussed
  in detail elsewhere~\cite{x11},
  there are several problems of internal consistency, particularly with
  tau-related measurements. The dashed line in Fig. 2 shows the weighted
  average value of $A_l$ if the \tpol~ measurements are excluded. It changes 
  by more than one standard deviation, and the internal consistency of the
  remaining data points is improved ($\chi^2 = 3.85$ for 3 DOF , CL =  28$\%$). 
    Repeating the analysis described above, but excluding the \tpol~ data,
  gives b quark couplings with even larger deviations ($\simeq 4\sigma$) from
  the SM predictions~\cite{x11}. Taking into account the CL (8.4$\%$) for self-
  consistency of the different $A_l$ measurements, the CL that all six effective
  couplings are consistent with lepton universality and the SM is 0.9$\%$.
  The similar CL for the leptonic and b quark couplings alone is 0.18$\%$.
    \par Also shown in Fig. 2 is the value of $A_l$ derived from
$A_{FB}^{0,b}$, assuming the correctness of the SM. The value so obtained,
0.1396(33), differs from the weighted average of the purely leptonic
 measurements by 2.6$\sigma$ It is clear, from the analysis presented
 above, that this discrepancy is 
 essentially due to the apparent deviations of the
 measured b quark effective
 couplings from the SM predictions. The quantity $\sin^2\theta_{eff}^{lept}$
 used in Ref.[1] is directly related to $A_l$ via Eqns.(4),(12). The poor
 consistency of the different $\sin^2\theta_{eff}^{lept}$ determinations
 in Table 19 of Ref.[1] results from the inclusion of values derived 
 from $A_{FB}^{0,b}$ and  $\langle A_{FB}^q \rangle$ which assume the
  correctness of the SM. This common 
 origin, in the measured b quark couplings, of the poor agreement of the
  different
  $\sin^2\theta_{eff}^{lept}$ determinations and the 3$\sigma$ deviation of
  the measured LEP-SLD average value of $A_b$ from the 
  SM prediction, was not pointed out in
  Ref.[1].
  
\section{Standard Model Comparison and Physical Interpretation}
 The test of the SM provided by measurements of Z decays
at LEP and SLD is, essentially, that of its 
predictions for the quantum corrections, arising from 
massive virtual particle loops, to the Born level diagrams for $e^+e^- \rightarrow
 Z \rightarrow f \overline{f}$.
The corrections may be conveniently expressed in terms of two parameters
$\Delta \rho_f$ and $\Delta \kappa_f$ for each fermion flavour~\cite{x12}.
 The parameters
 are given, in terms of the effective couplings, by the relations:
\begin{eqnarray}
\Delta \rho_f & = & -2(1-2|\abf|)  \\
\Delta \kappa_l & = & \frac{(1-\rbl)}{4 s_W^2}-1  \\
\Delta \kappa_c & = & 3\frac{(1-\rbc)}{8 s_W^2}-1  \\
\Delta \kappa_b & = & 3\frac{(1-\rbb)}{4 s_W^2}-1
\end{eqnarray}
Here, following the usual on-shell definition~\cite{x13}:
\begin{equation}
\sin^2 \theta_W = s_W^2 = 1-c_W^2 \equiv 1- \frac{M_W^2}{M_Z^2}
\end{equation}
\par The SM predictions of Section 2 used the fixed values:
$m_t = $ 172 GeV, $m_H = $ 149 GeV
found in the global fit of Ref.[1].
The effect on the SM prediction of varying $m_t$ and $m_H$ within the
existing experimental bounds~\cite{x1,x14} is now considered. The dependence
of $\Delta \rho_f$ on $m_t$ and $m_H$ is contained in the terms~\cite{x12}:
\begin{eqnarray}
\Delta \rho_f^{top} & = & \frac{ 3 G_{\mu} m_t^2}{8 \sqrt{2}\pi^2}(1+\xi_f) \\
\Delta \rho_f^{Higgs} & = &    -\frac{\sqrt{2} G_{\mu} M_W^2}{8 \pi^2}
\tan^2 \theta_W
\left[ \frac{11}{3}\left(\ln \left(\frac{m_H}{M_W}\right)-\frac{5}{12}\right)\right] 
\end{eqnarray}
where $\xi_f = 0$ for $f \ne b$ and -4/3 for  $f = b$. The quantum correction
$\Delta \kappa_f$ is calculated using a parameterisation\footnote{The 
relative accuracy of the formula (24) is about one per mille for the
interesting range of values of  $m_t$ and $m_H$.} of the ZFITTER~\cite{x15}
 prediction of the effective leptonic weak mixing angle:
\begin{equation}
(\overline{s}_W^l)^2 = 0.233597-8.95 \times 10^{-8}m_t^2-3.86 \times 10^{-4}
\ln m_t+5.43 \times 10^{-4}\ln m_H
\end{equation}
where  $m_t$ and $m_H$ are in GeV units. $\Delta \kappa_f$ is related to
$(\overline{s}_W^l)^2 = (1-\rbl)/4$ by Eqns. (1),(9),(10),(12) and (18-20).
 For the b quark there is an additional non-universal contribution:
\begin{equation}
\Delta \kappa_b^{top}  =  \frac{  G_{\mu} m_t^2}{4 \sqrt{2}\pi^2}.
\end{equation}
Although a reasonably good agreement is found for the leptonic and b quark
couplings with $m_t =$ 172 GeV and $m_H =$ 149 GeV somewhat better overall 
agreement is found for the choice $m_t =$ 180 GeV and $m_H =$ 100 GeV,
 still well within the current experimental
bounds~\cite{x1,x14}. The SM predictions for these values of $m_t$ and $m_H$ 
are also presented in Table 10. 
 For b quarks however,
the measured values of the quantum corrections are much larger than the SM
predictions. For $\Delta \rho_b$ the measured value exceeds the SM prediction
by a factor of 13, and is of opposite sign. The measured value of $\Delta \kappa_b$
has the same sign as the SM prediction, but is 9 times larger. Both effects are at
the $>$ 3 standard deviation level, but they are highly correlated. 
The discrepancies seen are
so large that the significance of the deviations  is almost independent of
the values of $m_t$ and $m_H$ assumed in the SM predictions.      
\par It is also instructive to present the quantum corrections in terms of the
`epsilon parameters' introduced by Altarelli et al.~\cite{x16,x17,x18}. In terms of 
the variables used in the present paper to describe the effective couplings, these
are defined as~\cite{x16}:
\begin{eqnarray}
\epsilon_1 & \equiv & \Delta \rho_l = -2(1+2\abl)  \\
\epsilon_2 & \equiv & c_0^2 \Delta \rho_l+
\frac{s_0^2 \Delta r_W }{(c_0^2-s_0^2)}-2 s_0^2 \Delta k' \\
\epsilon_3 & \equiv & c_0^2 \Delta \rho_l+(c_0^2-s_0^2) \Delta k'
\end{eqnarray}
here $s_0^2 = 1-c_0^2$ and $\Delta r_W$ are defined by the relations:
\[ \left(1-\frac{M_W^2}{M_Z^2}\right)\frac{M_W^2}{M_Z^2} = \frac{s_0^2 c_0^2}{1-\Delta r_W}
= \frac{\pi \alpha(M_Z)}{\sqrt{2} G_{\mu} M_Z^2(1-\Delta r_W)} \]
and 
\[ \Delta k' = \frac{(1-\rbl)}{4 s_0^2} -1. \]
In Ref.[18] a fourth parameter, $\epsilon_b$, was introduced. It may be
 defined in three distinct ways:
\begin{eqnarray} 
\epsilon_b(\abb) & \equiv & \frac{\abb}{\abl}-1   \\
\epsilon_b(\rbb) & \equiv & \frac{\rbb-{\cal R}_l}{1-\rbb}  \\
\epsilon_b(\sbb) & \equiv & \frac{\sbb-(\abl)^2(1-6 \mu_b+{\cal R}_l^2)}
{2 (\abl)^2 (1- 6 \mu_b+2{\cal R}_l)}
\end{eqnarray}
where
\[ {\cal R}_l \equiv \frac{(2+\rbl)}{3}. \]
In the SM, retaining only the leading terms $\simeq  m_t^2$, the three
definitions (29-31) are equivalent\footnote{ Modulo small b-mass
dependent corrections}. In previous phenomenological applications
however,~\cite{x18,x19} only the third definition (31) based, via Eqn.(8)
on the measurement of $R_b$ was used. The measured values of the
six epsilon parameters defined above are presented in Table 11, where 
they are compared with the SM predictions for the same values of $m_t$ and $m_H$
as in Table 10. As prevously noted~\cite{x19} the values of
$\epsilon_1$, $\epsilon_2$ and $\epsilon_3$ are in reasonably good agreement with
the SM predictions, with deviations in
the range 1-2 $\sigma$ . A small deviation is also observed for $\epsilon_b(\sbb)$, a
residual of the much discussed~\cite{x1} `$R_b$ problem'. However,
both $\epsilon_b(\abb)$ and $\epsilon_b(\rbb)$ deviate from the SM
prediction by about 4 standard deviations\footnote{Again, the errors
on these quantities are highly correlated}.
The parameter $\epsilon_b(\rbb)$ is very sensitive 
 to the anomalous b coupling; the measured value is 39 times and 
 4.5$\sigma$ larger than the
 SM prediction.\footnote{The errors on this quantity, determined essentially
by those on $A_b$, are skewed and non-gaussian. The average error is quoted
in Table 11. The confidence level of the deviation of
 $\epsilon_b(\rbb)$ from the SM, assuming gaussian errors for $A_b$,
 is in fact almost the same as that of the latter, about one per mille.}
 The SM predictions for these quantities are very insensitive to
 $m_H$ and are dominated
by the term $\simeq m_t^2$:
\begin{equation}
   \epsilon_b =   -\frac{G_{\mu} m_t^2}
   {4\sqrt{2} \pi^2} = -0.0062~~~(m_t = 172 {\rm GeV}).
\end{equation}.
\par The conclusion to be drawn from Table 11 is that the deviations 
observed for the b quark couplings, interpreted as a real physical 
effect, do not enter at all into the framework of the SM nor any of
its `natural' extensions. Supersymmetry, Technicolour, anomalous
$WW\gamma$ or $WWZ$ couplings, and new $U(1)$ gauge bosons are all
expected, via vacuum polarisation effects in the gauge boson 
propagators, to produce deviations from the SM predictions for $\epsilon_1$, $\epsilon_2$
 or $\epsilon_3$~\cite{x17}. No strong evidence for such deviations are observed, but rather 
 quantum corrections to the b quark
 couplings that disagree, by an order of magnitude, with the expectations of the SM.
 \par An indication of the possible 
 physical origin of the anomalous b quark
  couplings is provided by considering the right- and left-handed 
  effective couplings,
  $\gbr$, $\gbl$ related to
 $\abb$ and $\vbb$ by the relations:
\begin{eqnarray}
\gbr = \frac{1}{2}(\vbb-\abb) & = & -\sqrt{\rho_b} e_b (\overline{s}_W^b)^2 \\
\gbl = \frac{1}{2}(\vbb+\abb) & = &
 \sqrt{\rho_b}[T_b^3- e_b (\overline{s}_W^b)^2 ]
\end{eqnarray}
From the measured values of $\abb$ and $\vbb$ presented in Table 7, the following  
values of the left-handed and right-handed effective couplings of the b quarks are
found:
\[ \gbl  = -0.4155(30)~~~~~~\gbr  =  0.1098(101)  \]
which may be compared with the SM predictions of:
\[ \gbl  = -0.4208~~~~~~\gbr  =  0.0774  \]
The value of $\gbl$ is quite consistent with the SM prediction
 for $m_t = 172$ GeV, $m_H = 149$ GeV  
( a 1.3$\%$, 1.8$\sigma$ deviation) whereas the discrepancy for $\gbr$ is much larger,
(a 42$\%$, 3.2$\sigma$ deviation). One may remark that the weak isospin of the SM 
affects only $\gbl$, not $\gbr$, so it is possible that the SM does correctly describe
$\gbl$ but that there is a new, anomalous, right handed coupling for the b quark.
\par The right- and left-handed effective couplings of the s,d quarks have recently been 
measured by the OPAL collaboration\cite{x20} with the results: 
\[ \overline{g}^L_{d,s} = -0.44_{-0.09}^{+0.13}
~~~~~~\overline{g}^R_{d,s} = 0.13_{-0.17}^{+0.15} \]
to be compared with the SM predictions -0.424 and 0.077 respectively. These
 measurements are
in good agreement with both the SM predictions and the measured b quark
couplings given above.
\par Limits can also be set on possible anomalous couplings of the
other `d-type' quarks, d,s by comparing the measured values of
$\langle A_{FB}^q \rangle$ and $\Gamma_{had}$ with a model in which the 
d and s quarks are assumed to have the same effective coupling constants
as those measured for the b quarks. The prediction of this model for
$\langle A_{FB}^q \rangle $ is 0.1600(72), which is consistent with the
`measured' value (see Section 2 above and Table 8) of 0.1592(86) at the 
0.68$\sigma$ level. No useful constraint on possible anomalous couplings
of the d and s quarks, of a size similar to those observed for the b quark,
is therefore obtained using $\langle A_{FB}^q \rangle $ with the present
experimental errors. A more favourable case is  $\Gamma_{had}$.
Using the world average value of $\alps$ of 0.118(5)~\cite{x9,x10} in Eqn.(15)
to calculate the QCD correction, and with $<C_q^{QED}> = 1.00040$, the predicted
value of $\Gamma_{had}$ in the model with a universal right-handed anomaly
for down-type quarks is 1.7249(46) GeV. This differs by 3.6$\sigma$ from
the LEP average measurement~\cite{x1} of 1.7436(25) GeV. Thus this model is 
essentially excluded by the measurement of $\Gamma_{had}$.
It is interesting to note that the precise measurement of $\Gamma_{had}$
gives a more stringent constraint on possible anomalous couplings of the
d and s quarks than the direct measurement of their left- and right-handed
couplings cited above~\cite{x20}.
\par The values of the left- and right-handed couplings of the c quarks,
derived from the measured values of $\abc$ and $\vbc$ given in Table 7, are:  
\[ \gcl  = 0.3440(92)~~~~~~\gcr  =  -0.1600(70)  \]
in very good agreement with the SM predictions
for $m_t = 172$ GeV, $m_H = 149$ GeV of:
\[ \gcl  = 0.3465~~~~~~\gcr  =  -0.1545.  \]
The $\pm 2 \sigma$ limits for deviations of $\gcr$ from the SM prediction
extends from -0.174 to -0.146. Thus, at 95$\%$ CL, any anomalous 
right-handed couplings of the c quark lie between -9$\%$ and +15$\%$ of the
SM prediction.

\section{Summary}
The confidence level that all six effective couplings extracted in the
above analysis are consistent with lepton universality and the SM is
found to be 0.9$\%$. This number is the product of the CL for agreement
with lepton universality of the different $A_l$ measurements (8.4$\%$)
shown in Fig. 2
and the CL of the SM comparison using average quantites, shown in Table 9
(10.5$\%$). The similarly calculated CL that the leptonic and b quark 
couplings alone are consistent with lepton universality and the SM is
0.18$\%$. The measured effective couplings of the leptons and c quarks
agree well with the SM predictions, for top quark and Higgs boson masses
that are consistent with the current experimental limits.
 In contrast to this good agreement,
the extracted b quark couplings are found to deviate from the SM by
more than three standard deviations. The `epsilon parameters'
 extracted from the data show that such behaviour is
 not expected in any `natural' extension of the SM.
On the other hand, the parameter $\epsilon_b(\rbb)$ is found to be
particularly sensitive to anomalous behaviour of the b quark couplings;
the measured value differs from the SM prediction by 4.5$\sigma$.  
 Comparing the left
 and right-handed 
b quark couplings with the SM predictions, the former is consistent
below the 2$\sigma$ level
 whereas the latter shows a 42$\%$, or 3.2$\sigma$, deviation
from the prediction. Thus the only significant deviation observed
from the SM predictions for the effective couplings is due to the
 right-handed coupling of the b quark. 
\par Finally, it may be recalled that in 1995 a difference in the
world average value of $R_b$ from the
SM prediction of about three standard deviations was reported~\cite{x1}.
 The most recent average value is in much better agreement with the SM
prediction, due to a better understanding of experimental systematics.
However, as the $A_{FB}^{0,b}$ value quoted in Table 1 has a statistical
error very close to the expected final LEP1 value, no large change in the
average value is to be expected in the future. Of course, a hitherto
unknown and correlated systematic effect in all the LEP $A_{FB}^{0,b}$
measurements cannot be excluded as the source of the apparent deviation
from the SM seen in $\rbb$. Unfortunately, the direct measurement
of $A_b$ at SLD is not expected to be of sufficient precision, even
at the end of the experimental program, to shed much light on this
possiblity. A more detailed statistical discussion of the expected future
improvements in electroweak measurements, and of their probable effects
on the value of $\rbb$, can be found in Reference 11.

\section{Acknowledgements}
I thank M.Consoli for reading an early draft and for a
useful discussion. I am also grateful to J.Mnich for providing me with 
results from the ZFITTER program that revealed an error in
the Standard Model predictions in the previous version of this
letter. I thank the referee for remarks that have improved
 the clarity of the presentation, and for suggesting a  
detailed investigation of the sensitivity of the deviations observed
for the b quark couplings on the treatment of statistical and systematic
errors.  Finally, I should
like to thank T.Junk for first making me aware of the potential problems
 posed, for the Standard Electroweak Model, by the measurements of the right 
handed b quark coupling.

\pagebreak

\pagebreak
\begin{table}
\begin{center}
\begin{tabular}{|c|c|c|c|} \hline
  Quantity  &  Value ($\sigma_{TOT}$) [$\sigma_{SYS}$]   & SM &
 (Meas.-SM)/$\sigma_{TOT}$ \\
\hline  
  LEP        &   &   &    \\  \cline{1-1}
  $A_{FB}^{0,e}$   & 0.0160(24) [16] & 0.0159 & 0.04  \\
  $A_{FB}^{0,\mu}$   & 0.0162(13) [5] & 0.0159 & 0.04  \\
  $A_{FB}^{0,\tau}$   & 0.0201(18) [10] & 0.0159 & 2.3  \\ 
  $\Gamma_l$ (Mev)   & 83.91(11) [8] & 83.96 & -0.45  \\ 
  \tpol     &  &  & \\
  $A_e$  &   0.1382(76) [21] & 0.1458 & -1.0 \\
  $A_{\tau}$  &   0.1401(67) [45] &  0.1458 & -0.9 \\
  c and b quarks  &  &  &  \\
   $A_{FB}^{0,c}$   & 0.0733(49) [26] & 0.0730 &  0.1   \\ 
   $R_c$            & 0.1715(56) [42]  & 0.1723 & -0.1 \\
   $A_{FB}^{0,b}$   & 0.0979(23) [10]  & 0.1022 & -1.8  \\ 
   $R_b$            & 0.2179(12) [9]  & 0.2158 &  1.8 \\ 
\hline 
  SLD   &  &  & \\ \cline{1-1}
  $A_e$  &   0.1543(37) [14]  & 0.1458 & 2.3 \\
  $A_c$  &   0.625(84) [41] & 0.667 & -0.5 \\
  $A_b$  &   0.863(49) [32] & 0.935 & -1.4 \\
  $R_b$  &   0.2149(38) [21] & 0.2158 & -0.2 \\              
\hline
\end{tabular}
\caption[]{ Average values of electroweak observables used
 in the analysis~\cite{x1}. SM denotes the Standard Model
 prediction for $m_t =$ 172 GeV, 
$m_H =$ 149 GeV~\cite{x1}. $\sigma_{TOT}$ is the total experimental
error, which is the quadratic sum of the statistical ($\sigma_{STAT}$) and
systematic ($\sigma_{SYS}$) errors.   }
\end{center}
\end{table} 
\begin{table}
\begin{center}
\begin{tabular}{|c|c|c|c|c|} \hline
  $A_l$  & $A_c$  & $A_b$  & $R_c$ &  $R_b$  \\  
\hline
0.1501(24) & 0.645(39) & 0.869(22) & 0.1715(56) & 0.2176(11) \\
\hline
\end{tabular}
\caption[]{ LEP+SLD averages}  
\end{center}
\end{table}
\begin{table}
\begin{center}
\begin{tabular}{|c|c|c|c|} \hline
      & $\rbl$  & $\rbc$  &  $\rbb$   \\
 \hline      
Measurement &  0.07548(120) &  0.366(29)   & 0.582(32)  \\
     SM &  0.07332 &  0.383   & 0.689  \\
 (Meas.-SM)/Error&  -1.80  & -0.59 & -3.34  \\  
\hline
\end{tabular}
\caption[]{ Measured values of $\rbf = \vbf/\abf$ compared to Standard Model 
predictions } 
\end{center}
\end{table}
\begin{table}
\begin{center}
\begin{tabular}{|c|c|c|c|} \hline
      & $\sbl$  & $\sbc$  &  $\sbb$   \\
\hline      
Measurement &  0.25244(33) &  0.2877(95)   & 0.3676(24)  \\
     SM &  0.25259 &  0.2880   & 0.3644  \\
 (Meas.-SM)/Error &  -0.45  & -0.03 & 1.33  \\
\hline
\end{tabular}
\caption[]{ Measured values of $\sbf = \abf^2(1-6 \mu_f)+\vbf^2$  compared 
to Standard Model predictions } 
\end{center}
\end{table}
\begin{table}
\begin{center}
\begin{tabular}{|c|c|c|c|} \hline
      & \multicolumn{3}{c|}{leptons}  \\ \cline{2-4}       
      & Meas. & SM & Dev($\sigma$)  \\
\hline  
$\abf$ & -0.50101(33) & -0.50124 &  0.67 \\ 
$\vbf$ & -0.03782(68) & -0.03675 &  -1.57 \\              
 \hline
\end{tabular}
\caption[]{ Measured values of the effective electroweak coupling 
constants for the charged leptons. Dev($\sigma$) = (Meas.-SM)/Error. }
\end{center}
\end{table}
\begin{table}
\begin{center}
\begin{tabular}{|c|c|c|c|} \hline
  $C_c^{QED}$ & $C_b^{QED}$  & $C_c^{QCD}$  & $C_b^{QCD}$   \\
\hline      
 1.00046 &  0.99975 &  1.0002   & 0.9953  \\
\hline
\end{tabular}
\caption[]{ QED and QCD correction factors for heavy quarks
assuming $\alpha_s(M_Z) = 0.12$ and $\alpha(M_Z)^{-1} = 128.9$.  } 
\end{center}
\end{table}
\begin{table}
\begin{center}
\begin{tabular}{|c|c|c|c|c|c|c|} \hline
      & \multicolumn{3}{c|}{c quark} 
      &  \multicolumn{3}{c|}{b quark} \\  \cline{2-7}    
      & Meas. & SM & Dev($\sigma$)
      & Meas. & SM & Dev($\sigma$)  \\
\hline  
$\abf$ & 0.504(10) & 0.501 & 0.30 
       & -0.5252(75) & -0.4981 & -3.61  \\ 
$\vbf$ & 0.184(15) & 0.192 & -0.53 
       & -0.3057(125) & -0.3434 & 3.18  \\              
 \hline
\end{tabular}
\caption[]{ Measured values of the effective electroweak coupling 
constants of c and b quarks. Dev($\sigma$) = (Meas.-SM)/Error. } 
\end{center}
\end{table}
\begin{table}
\begin{center}
\begin{tabular}{|c|c|c|c|c|} \hline
   &`Measured' &  SM Pred. & MI Pred.  & MI Pred. with SM b quark \\
\hline      
 $\langle A_{FB}^q \rangle$  & 0.1592(86) & 0.1641 & 0.1639(28) & 
 0.1692(28)  \\
\hline
(`Meas'.-Pred.)/Error &  & -0.57 & -0.52  & -1.1 \\
\hline
\end{tabular}
\caption[]{ Values of the mean quark charge asymmetry. `MI Pred.'
stands for Model Independent Prediction (see text). See also
the text for the definition of `Measured'.   }
\end{center}
\end{table} 
\begin{table}
\begin{center}
\begin{tabular}{|c|c|c|c|c|c|} \hline
  Observables & $\rbl$, $\sbl$ & $\rbl$, $\rbb$  &
   $\rbl$, $\sbl$, $\rbb$, $\sbb$ & $\rbl$, $\rbb$, $\rbc$ &
   $\rbl$, $\sbl$, $\rbb$, $\sbb$, $\rbc$, $\sbc$  \\
\hline  
 $dof$ & 2 & 3 & 5 & 5 & 8 \\
 $\chi^2$ & 3.44 & 10.6 & 13.2 & 10.9 & 13.2 \\
 CL ($\%$) & 17.9 & 1.4 & 2.2 & 5.3 & 10.5 \\
\hline 
\end{tabular}
\caption[]{ $\chi^2$ and confidence levels for agreement with the SM
( $m_t =$ 172 GeV, $m_H =$ 149 GeV) of different
 sets of electroweak observables sensitive to the
effective couplings, assuming perfect statistical consistency
 of the LEP+SLD averages in Table 2. See the text for the explanation of
 the number of degrees of freedom ($dof$) in each case.} 
\end{center}
\end{table}
\begin{table}
\begin{center}
\begin{tabular}{|c|c|c|c|c|c|c|} \hline
      & \multicolumn{2}{c|}{leptons} 
      & \multicolumn{2}{c|}{c quark}
      & \multicolumn{2}{c|}{b quark} \\  \cline{2-7}    
      & $\Delta \rho_l$ & $\Delta \kappa_l$ &  $\Delta \rho_c$
      & $\Delta \kappa_c$ & $\Delta \rho_b$ & $\Delta \kappa_b$  \\
\hline  
Expt. & 0.00404(133) & 0.03445(134) & 0.016(41) 
       & 0.064(49) & 0.101(32) & 0.403(107)  \\
\hline
SM $m_t =$ 172 GeV&   &   &   &   &   & \\         
 $m_H =$ 149 GeV & 0.00497 & 0.03686 & 0.005 
       & 0.037 & -0.007 & 0.0436  \\
 Dev($\sigma$) & -0.7 & -1.8 & 0.27 & 0.55 & 3.38 & 3.36 \\                     
 \hline
SM $m_t =$ 180 GeV&   &   &   &   &   & \\     
   $m_H =$ 100 GeV & 0.00563 & 0.03472 & 0.006 
       & 0.034 & -0.008 & 0.0412  \\
 Dev($\sigma$) & -1.2 & -0.02 & 0.24 & 0.61 & 3.40 & 3.38 \\                     
 \hline 
\end{tabular}
\caption[]{ Measured values of the quantum correction parameters
$\Delta \rho_f$ and $\Delta \kappa_f$ compared to SM predictions.
 Dev($\sigma$) = (Expt.-SM)/Error. }
\end{center}
\end{table}
\begin{table}
\begin{center}
\begin{tabular}{|c|c|c|c|c|c|c|} \hline
  & $\epsilon_1$ & $\epsilon_2$ & $\epsilon_3$ & $\epsilon_b(\sbb)$
  & $\epsilon_b(\abb)$ & $\epsilon_b(\rbb)$ \\
\hline  
Expt. & 0.00404(133) & -0.0073(8) & 0.0031(8) 
       & -0.0017(18) & 0.048(15) & -0.263(57)  \\
\hline
SM $m_t =$ 172 GeV&   &   &   &   &   & \\         
   $m_H =$ 149 GeV & 0.00497 & -0.0076 & 0.0051 
       & -0.0045 & -0.0060 & -0.0068  \\
 Dev($\sigma$) & -0.7 & 0.38 & -2.5 & 1.6 & 3.6 & -4.5 \\                     
 \hline
SM $m_t =$ 180 GeV&   &   &   &   &   & \\  
   $m_H =$ 100 GeV & 0.00563 & -0.0062 & 0.0045 
       & -0.0046 & -0.0068 & -0.0055  \\
 Dev($\sigma$) & -1.2 & -1.4 & -1.8 & 1.6 & 3.7 & -4.5 \\                     
 \hline 
\end{tabular}
\caption[]{ Measured values of the epsilon parameters of 
Refs.[16-18] compared to SM predictions.
 Dev($\sigma$) = (Expt.-SM)/Error. }
\end{center}
\end{table}

\end{document}